# Reversible Reciprocal Relation of Thermoelectricity

Yu-Chao Hua, Ti-Wei Xue, and Zeng-Yuan Guo[*]

Key Laboratory for Thermal Science and Power Engineering of Ministry of Education, Department of Engineering Mechanics, Tsinghua University, Beijing 100084, P. R. China

[*] Corresponding author: E-mail: demgzy@tsinghua.edu.cn

**ABSTRACT:**

The first Kelvin relation that states the Peltier coefficient should be equal to the product of temperature and Seebeck coefficient is a fundamental principle in thermoelectricity. It has been regarded as an important application and direct experimental verification of Onsager reciprocal relation (ORR) that is a cornerstone of irreversible thermodynamics. However, some critical questions still remain: (a) why Kelvin's proof that omits all irreversibility within a thermoelectric transport process can reach the correct result, (b) how to properly select the generalized-force-flux pairs for deriving the first Kelvin relation from ORR, and (c) whether the first Kelvin relation is restricted by the requirement of linear transport regime. The present work is to answer these questions based on the fundamental thermodynamic principles. Since the thermoelectric effects are reversible, we can redefine the Seebeck and Peltier coefficients using the quantities in reversible processes with no time derivative involved, which are renamed as "reversible Seebeck and Peltier coefficients". The relation between them (called "the reversible reciprocal relation of thermoelectricity") is derived from the Maxwell relations, which can be reduced to the conventional Kelvin relation, when the local equilibrium assumption (LEA) is adopted. In this sense, the validity of the first Kelvin relation is guaranteed by the reversible thermodynamic principles and LEA, without the requirement of linear transport process. Additionally, the generalized force-flux pairs to obtain the first Kelvin relation from ORR can be proper both mathematically and thermodynamically, only when they correspond to the conjugate-variable pairs of which Maxwell relations can yield the reversible reciprocal relation. The present theoretical framework can be further extended to coupled phenomena.

# 1. Introduction

Thermoelectric effects refer to the direct conversion of heat to electric energy and vice versa, which encompasses three separately identified phenomena [1]: the Seebeck effect, Peltier effect, and Thomson effect, with three corresponding coefficients. The Seebeck coefficient ($\alpha$) is the ratio between the temperature-gradient-induced voltage gradient and the temperature gradient, the Peltier coefficient ($\Pi$) is defined as the ratio between the electric-current-induced heat flow and the electric current, and the Thomson coefficient ($\chi$) is the ratio between the heat production rate per unit volume and the product of electric current and temperature gradient. All the three thermoelectric effects are thermodynamically reversible [2], which means the inversion of the direction of driving force (or flux) will cause the inversion of the corresponding effect. By contrast, the thermoelectric coefficients are conventionally defined using the quantities of irreversible transport processes (such as, electric current and temperature gradient) with time derivatives involved.

Early in 1851, Kelvin (also known as Thomson) [3] identified that these three effects are not independent from each other, and proposed the Kelvin relations (also known as Thomson relations) that correlate the Seebcek, Peltier and Thomson coefficients on the basis of fundamental reversible thermodynamics [4]. The first Kelvin relation is $\alpha T = \Pi$ with temperature $T$, while the second Kelvin relation gives $\chi = T\, d\alpha/dT$. In 1893, the Kelvin relations were first verified experimentally, and from then on they have been widely accepted and utilized [5, 6]. It is noted that the second Kelvin relation can be readily derived from the first Kelvin relation, energy conservation law, and local equilibrium assumption [1]; in this sense, the first Kelvin relation should be the core of the relations among the three thermoelectric effects.

To derive the relations, Kelvin assumed a short circuit comprised of two kinds of materials with two contact junctions that are in contact with heat reservoirs of different temperatures; all irreversible factors, including Joule heating and heat conduction, are neglected, and the conservation of energy and entropy flows will yield the required relations. Nevertheless, Kelvin's proof is questionable indeed [7], since these irreversible factors should not be neglected in the circuit with a finite temperature difference. As stated by Onsager in his famous paper in 1931 [7], "*Thomson's relation has not been derived entirely from recognized fundamental principles, nor is it known exactly which general laws of molecular mechanics might be responsible for the success of Thomson's peculiar hypothesis.*" Even at now, the question has not been answered satisfactorily why Kelvin's proof that omits all irreversible factors within a thermoelectric transport process can reach the correct result.

In 1931, Onsager [7, 8] developed the Onsager reciprocal relations (ORR) on the basis of statistical thermodynamics and microscopic reversibility. The first Kelvin relation can be readily derived from ORR when the generalized-force-flux pairs are chosen properly [9, 10]. Afterwards, the Kelvin relation has been put on a solid physical basis, and reciprocally its experiments have been regarded as an important experimental verification of ORR [5, 11]. However, when taking a close look at the derivation from the ORR to the first Kelvin relation, one critical issue exists, that is how to properly select the generalized-force-flux pairs to construct the linear phenomenological relations. Some researchers summarized two requirements [10] on the selection of generalized fluxes $J$ and forces $X$: (i) the product of $J$ and $X$ is equal to the local entropy production rate $\sigma_S$, that is, $\sigma_S = J \cdot X$ (R1), and (ii) $J$ is the time derivative of a *state variable*, and $X$ is the derivative of the entropy deviation with respect to the state variable (R2). However, counter examples do exist. For instance, in Miller's paper [5], the generalized flux and force are heat flux $q_h$ and $-\nabla T/T$ for heat conduction, while the charge transport's generalized flux and force are electric current $I_e$ and the negative voltage gradient $-\nabla V_e$; the products of generalized force and flux are $T\sigma_S$ rather than $\sigma_S$.

Additionally, the ORR is based on the linear phenomenological relations between generalized forces and fluxes [12, 13], which indicates that the first Kelvin relation should be valid merely in the regimes of linear transport. However, in practice the first Kelvin relation is generally employed to handle the problems with large temperature difference where the linearity of constitutive relations could violate [14]. Therefore, it is also needed to clarify whether the first Kelvin relation is restricted by the requirement of linearity.

The present work is trying to answer the questions above on the basis of fundamental thermodynamic principles. Firstly, we will introduce the concepts of reversible Seebeck and Peltier coefficients with no time derivative involved, and clarify their relations with the conventional thermoelectric coefficients. Then, the relation between reversible Seebeck and Peltier coefficients will be derived from the Maxwell relations. Based on it, the selection of generalized-flux-force pairs and the requirement of linearity are clarified.

## 2. Reversible Seebeck and Peltier coefficients

The thermoelectric effects are reversible, since the inversion of the direction of driving force (or flux) can inverse the resulted effect and theoretically the efficiency of a thermoelectric generator is approaching Carnot efficiency in a limiting case [2]; thus, they can be analyzed in terms of the fundamental reversible thermodynamics. To do this, we redefine the thermoelectric coefficients using the quantities in the reversible processes with no time derivative involved, and for convenience we call them "reversible Seebeck and Peltier coefficients" Note that we here focus on the thermoelectric effects in solids with volume kept constant.

Figure 1 shows an electrically-insulated subsystem that is embedded in an outer system with infinite thermal and electric capacitances. In this combined system, the temperature of outer system is gradually increased by an infinitesimal temperature change $\Delta T$, with the iso-electrochemical-potential condition of outer system; a quasi-static reversible process occurs, during which heat is transferred from the outer system to the subsystem until the new equilibrium state of subsystem is reached. At this new equilibrium state, the subsystem's quantity of charge is unchanged due to its electrically-insulated boundaries, and instead the thermoelectric effect will cause a change of electrochemical potential, $\Delta\mu_e = e\Delta V_e$, with $\Delta V_e$ the voltage change and $e$ the elementary charge. In this case, a reversible Seebeck coefficient can be defined as,

$$\alpha_r = -\frac{\Delta V_e}{\Delta T} = -\frac{\Delta\mu_e}{e\Delta T}. \tag{1}$$

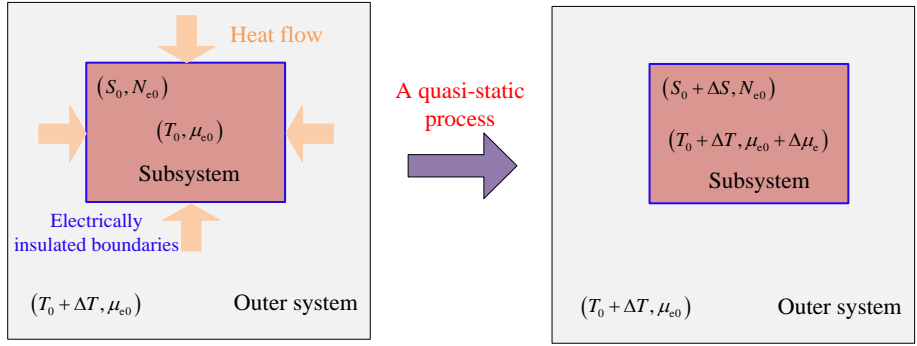

Figure 1 Schematics for the definition of reversible Seebeck coefficient: A quasi-static process driven by an infinitesimal $\Delta T$ with the iso-electrochemical-potential condition of outer system.

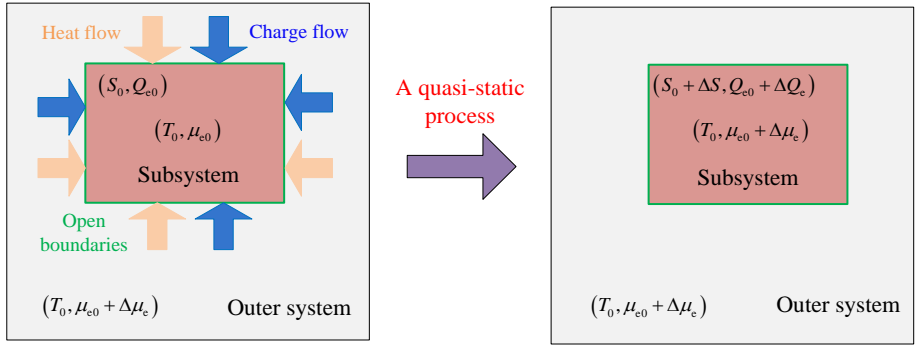

Figure 2 Schematics for the definition of reversible Peltier coefficient: A quasi-static process driven by an infinitesimal $\Delta\mu_e$ with the isothermal condition of outer system.

Similarly, a reversible Peltier coefficient is defined by constructing a quasi-static process shown in Fig. 2. A subsystem with open boundaries is embedded in the outer system. An infinitesimal change of electrochemical potential $\Delta\mu_e$ is introduced to the outer system with the isothermal condition; $\Delta\mu_e$ drives both heat and charge quantity to transfer from the outer system to the subsystem until the new equilibrium state of subsystem is reached. Finally, for the subsystem, the changes of entropy $S$ and charge quantity $Q_e$ are denoted by $\Delta S$ and $\Delta Q_e$ respectively. The change of heat $\Delta Q_h$ is equal to $T_0 \Delta S$ with $T_0$ the temperature kept constant during the quasi-static process. The change of charge quantity $\Delta Q_e$ should be equal to $e\Delta N_e$ where $N_e$ is the number of charge particles. Thus, the reversible Peltier coefficient is given by,

$$\Pi_r = \frac{\Delta Q_h}{\Delta Q_e} = \frac{T_0 \Delta S}{e \Delta N_e}, \qquad (2)$$

which is "heat per carrier" in the quasi-static reversible process.

## 3. Maxwell relations and the first Kelvin relation

*3.1 Derivation of the reversible reciprocal relation of thermoelectricity from Maxwell relations*

We start from the basic equations in reversible thermodynamics [15],

First law of thermodynamics:

$$TdS = dU + \sum_i -\mu_i dN_i + pdV, \qquad (3)$$

Euler integration:

$$ST = U + \sum_i -\mu_i N_i + pV, \qquad (4)$$

and Gibbs-Duhem equation:

$$SdT = \sum_i -N_i d\mu_i + Vdp. \qquad (5)$$

Setting $V = V_0 = \text{const}$ for solids, we have

$$dU = TdS + \sum_i \mu_i dN_i \qquad (6)$$

$$d(V_0 p) = V_0 dp = SdT + \sum_i N_i d\mu_i \qquad (7)$$

Particularly for thermoelectric systems, Eqs. (6) & (7) become

$$dU = TdS + \mu_e dN_e, \tag{8}$$

$$d(V_0 p) = SdT + N_e d\mu_e. \tag{9}$$

Two pairs of conjugate variables are selected in terms of Eqs. (8) & (9), which are

$$\{S, T\}, \{N_e, \mu_e\}, \tag{10}$$

where $S$ and $N_e$ are extensive variables, while $T$ and $\mu_e$ are intensive ones. Two exact differentials are thus given by,

$$-dS = \Gamma_{11} dT + \Gamma_{12} d\mu_e, \tag{11}$$

$$-dN_e = \Gamma_{21} dT + \Gamma_{22} d\mu_e. \tag{12}$$

Note that the parameters, $\Gamma_{11}$, $\Gamma_{12}$, $\Gamma_{21}$, and $\Gamma_{22}$, are not necessary to be constant. The combination of Eqs. (9), (11) &(12) leads to a Maxwell relation [15] of $\Gamma_{12}$ and $\Gamma_{21}$ (MR1),

$$\Gamma_{12} = \left.\frac{-dS}{d\mu_e}\right|_T = -\frac{d^2(V_0 p)}{dTd\mu_e} = -\frac{d^2(V_0 p)}{d\mu_e dT} = \left.\frac{-dN_e}{dT}\right|_{\mu_e} = \Gamma_{21}. \tag{13}$$

For the quasi-static process shown in Fig. 1 where $N_e$ is unchanged, Eq. (11) becomes,

$$dN_e = \Gamma_{21} dT + \Gamma_{22} d\mu_e = 0 \Rightarrow \Gamma_{21} \Delta T + \Gamma_{22} \Delta \mu_e = 0. \tag{14}$$

Referring to Eq. (1), the reversible Seebeck coefficient is given by,

$$-\frac{\Delta \mu_e}{e\Delta T} = \alpha_r = \frac{1}{e}\frac{\Gamma_{21}}{\Gamma_{22}}. \tag{15}$$

Furthermore, for the quasi-static process shown in Fig. 2 with $dT$ vanishing, Eqs. (11) & (12) are reduced to

$$-dS = \Gamma_{12} d\mu_e \Rightarrow -\Delta S = \Gamma_{12} \Delta \mu_e, \tag{16}$$

$$-dN_e = \Gamma_{22} d\mu_e \Rightarrow -\Delta N_e = \Gamma_{22} \Delta \mu_e. \tag{17}$$

Thus, according to Eq. (2), the reversible Peltier coefficient is,

$$\Pi_r = \frac{\Delta Q_h}{e\Delta N_e} = \frac{T}{e}\frac{\Gamma_{12}}{\Gamma_{22}}. \tag{18}$$

Apparently, in terms of MR1, we have

$$T\alpha_r = \Pi_r. \tag{19}$$

Actually, we can derive another two exact differentials for this conjugate-variable pairs, which are reciprocal to Eqs. (11) and (12),

$$\begin{aligned}-dT &= R_{11}dS + R_{12}dN_e,\\ -d\mu_e &= R_{21}dS + R_{22}dN_e.\end{aligned} \tag{20}$$

Then, combining Eqs. (8)&(20) yields a Maxwell relation of $R_{12}$ and $R_{21}$(MR2),

$$R_{12} = -\frac{dT}{dN_e}\bigg|_S = -\frac{d^2U}{dN_e dS} = -\frac{d^2U}{dSdN_e} = \frac{d\mu_e}{dS}\bigg|_{N_e} = -R_{21}. \tag{21}$$

Following the identical procedure, the same relation as Eq. (19) can be recovered from MR2.

The derivation above yields a relation that shows the reversible Peltier coefficient is equal to the product of temperature and reversible Seebeck coefficient, which has the same form as the conventional first Kelvin relation. For convenience, we call this relation "the reversible reciprocal relation of thermoelectricity".

*3.2 Relation between the conventional Kelvin relation and the reversible reciprocal relation of thermoelectricity*

In order to clarity the relation between the conventional Kelvin relation and the reversible reciprocal relation of thermoelectricity, we need to analyze the relationship between the conventional and reversible thermoelectric coefficients.

The conventional Seebeck and Peltier coefficients are defined in irreversible transport processes, which are given by [1],

$$\alpha = -\frac{\nabla V_e}{\nabla T}, \quad (22)$$

with temperature gradient $\nabla T$ and temperature-gradient-induced voltage gradient $\nabla V_e$, and

$$\Pi = \frac{\boldsymbol{q}_h}{\boldsymbol{I}_e}, \quad (23)$$

with heat flux $\boldsymbol{q}_h$ and electric current $\boldsymbol{I}_e$. In fact, based on the Kelvin relation, $\alpha T = \Pi$, the conventional Seebeck coefficient can be transformed to [9],

$$\alpha = \frac{\boldsymbol{q}_S}{\boldsymbol{I}_e}, \quad (24)$$

in which $\boldsymbol{q}_S$ is entropy flux. Therefore, in the irreversible thermoelectric transport processes, $\Pi$ is "heat per carrier", and $\alpha$ is "entropy per carrier", according to Eqs. (23) and (24).

Furthermore, according to Eq. (2), the reversible Peltier coefficient is "heat per carrier" in the reversible process. Importantly, using the reversible reciprocal relation of thermoelectricity, Eq. (19), the reversible Seebeck coefficient can be expressed as,

$$\alpha_r = \frac{\Delta S}{\Delta Q_e}, \quad (25)$$

which is "entropy per carrier" in the reversible process. Therefore, when the local equilibrium assumption is valid, the reversible thermoelectric coefficients should be equivalent to the conventional ones, and thus these two relations are also equivalent. This well explains why Kelvin's proof that omits all the irreversible factors can reach the correct result: Kelvin used the basic equations in reversible thermodynamics, and thus could obtain the reversible reciprocal relation of thermoelectricity.

In addition, the derivation above demonstrates that the Kelvin relations are not restricted by the requirement of linear phenomenological relations. In fact, the Kelvin relations will hold, once the local equilibrium assumption and the fundamental thermodynamic principles are valid.

*3.3 Selection of generalized force-flux pairs in the terms of conjugate-variable pairs*

Table 1 summarizes the generalized force-flux pairs in literatures [1, 5, 9]. Apparently, referring to Tab.1, the selection of generalized force-flux pairs to obtain the Kelvin relation from ORR is not unique nor arbitrary. The product of generalized forces and fluxes is either $\sigma_S$ or $T\sigma_S$. Note that heat flux $\boldsymbol{q}_h$ can be regarded as the time derivative of a state variable, merely when assuming the heat capacity is a function only dependent on temperature.

As a reasonable inference, the proper generalized force-flux pairs should correspond to the conjugate-variable pairs of which Maxwell relations can yield the reversible reciprocal relation of thermoelectricity. The conjugate-variable pairs, $\{S,T\},\{N_e,\mu_e\}$, is corresponding to the generalized force-flux pairs, $\{\boldsymbol{q}_S,-\nabla T\},\{\boldsymbol{I}_e,-\nabla\mu_e\}$, and the product of generalized force and flux is equal to $T\sigma_S$. Moreover, the reversible reciprocal relation of thermoelectricity can also be derived from another set of conjugate-variable pairs (the relevant proof is given in Appendix 1), and they are,

$$\left\{U,\frac{1}{T}\right\},\left\{N_e,\frac{-\mu_e}{T}\right\}, \tag{26}$$

which corresponds to the generalized force-flux pairs,

$\{\boldsymbol{q}_U, \nabla(1/T)\}, \{\boldsymbol{I}_e, \nabla(-\mu_e/eT)\}$ with total energy flux $\boldsymbol{q}_U$. In this case, the product of generalized force and flux becomes $\sigma_S$.

In this sense, no conjugate-variable pair can be constructed to correspond to the generalized force-flux pairs involving heat flux, since heat flux is not the time derivative of any state variable in thermodynamics. Thus, we need further clarify why these generalized force-flux pairs involving heat flux can also derive the first Kelvin relation.

Take $\{\boldsymbol{q}_h, -\nabla \ln T\}, \{\boldsymbol{I}_e, -\nabla(\mu_e/e)\}$ as an example. This set of generalized force-flux pairs implies such two "exact" differentials,

$$-\delta Q_h = \Gamma_{11}^* \frac{1}{T} dT + \Gamma_{12}^* d\mu_e, \tag{27}$$

$$-dN_e = \Gamma_{21}^* \frac{1}{T} dT + \Gamma_{22}^* d\mu_e. \tag{28}$$

Then, we have

$$\begin{aligned} \Gamma_{12}^* &= -\frac{\delta Q_h}{d\mu_e}\bigg|_T = -T \frac{dS}{d\mu_e}\bigg|_T \\ \Gamma_{21}^* &= -T \frac{\delta N_e}{dT}\bigg|_{\mu_e} \end{aligned}. \tag{29}$$

According to MR1, Eq. (13), we still have $\Gamma_{12}^* = \Gamma_{21}^*$, which indicates Eqs. (27) and (28) can also derive the reversible reciprocal relation of thermoelectricity. However, it is emphasized that Eq. (27) is definitely illegal in the sense of thermodynamics, since $Q_h$ is a process variable. Thus, the generalized force-flux pairs involving heat flux may be proper in the view of mathematics, but incorrect in the sense of thermodynamics.

Table 1 Generalized force-flux pairs for obtaining the first Kelvin relation from ORR.

| | | Force | Flux | Product of forces and fluxes |
|---|---|---|---|---|
| 1 | Thermal | $\nabla(1/T)$ | Total energy flux $q_U$ | $\sigma_S$ |
| | Electric | $\nabla(-\mu_e/eT)$ | $I_e$ | |
| 2 | Thermal | $\nabla(1/T)$ | $q_h$ | |
| | Electric | $\frac{1}{T}\nabla(-\mu_e/e)$ | $I_e$ | |
| 3 | Thermal | $-\nabla \ln T$ | $q_S$ | |
| | Electric | $\frac{1}{T}\nabla(-\mu_e/e)$ | $I_e$ | |
| 4 | Thermal | $-\nabla T/T$ | $q_h$ | $T\sigma_S$ |
| | Electric | $\nabla(-\mu_e/e)$ | $I_e$ | |
| 5 | Thermal | $-\nabla T$ | $q_S$ | |
| | Electric | $\nabla(-\mu_e/e)$ | $I_e$ | |

## 4. Conclusions

(1) To analyze the thermoelectric effects using the fundamental reversible thermodynamics, we redefine the Seebeck and Peltier coefficients using the quantities in reversible processes with no time derivative involved. Based on the Maxwell relation derived from the fundamental principles of equilibrium thermodynamics, we demonstrate that the relation between the reversible Seebeck and Peltier coefficients, i.e., the reversible reciprocal relation of thermoelectricity, has the same from as that between the conventional ones.

(2) When the local equilibrium assumption holds, the reversible thermoelectric coefficients should be equivalent to the conventional ones; in this case, the reversible reciprocal relation of thermoelectricity is reduced to the conventional first Kelvin relation.

(3) The first Kelvin relation should not be restricted by the requirement of linear phenomenological relations, and it will hold once the local equilibrium assumption and the fundamental thermodynamic principles are valid.

(4) Since Kelvin omitted all the irreversible factors in a thermoelectric transport process with finite temperature difference, what he derived from the fundamental balance equations in equilibrium thermodynamics is just the reversible reciprocal relation of thermoelectricity. This explains why the questionable proof by Kelvin can lead to the correct result.

(5) For obtaining the first Kelvin relation from ORR, the generalized force-flux pairs, which are proper in the sense of both mathematics and thermodynamics, should correspond to the conjugate-variable pairs of which Maxwell relations can yield the reversible reciprocal relation of thermoelectricity.

(6) Although the present theoretical framework is used to analyze the thermoelectric effects, it can be extended to deal with other types of coupled phenomena, such as electrokinetics and heat-moisture-coupled transport.

**Acknowledgements**

The authors would like to express the very great appreciation to Prof. Yuan Dong, Prof. Hai-Dong Wang, Prof. Bai Song, Dr. Tian Zhao, and Dr. Sheng-Zhi Xu for their valuable and constructive suggestions during the planning and development of this research work. This work is financially supported by National Natural Science

Foundation of China (No. 51825601, 51676108), the Initiative Postdocs Supporting Program of China Postdoctoral Science Foundation (No. BX20180155), Project funded by China Postdoctoral Science Foundation (No. 2018M641348), Science Fund for Creative Research Group (No. 51321002).

# Appendix 1

This section provides a proof on how to derive the reversible reciprocal relation of thermoelectricity from the conjugate-variable pairs,

$$\left\{U, \frac{1}{T}\right\}, \left\{N_e, \frac{-\mu_e}{T}\right\}.$$

For solid-state thermoelectric systems, we have,

First law of thermodynamics:

$$TdS = dU - \mu_e dN_e, \qquad (A1)$$

Euler integration:

$$ST = U - \mu_e dN_e + pV_0. \qquad (A2)$$

Eq. (A1) can be transformed to,

$$dS = \frac{1}{T}dU + \left(-\frac{\mu_e}{T}\right)dN_e. \qquad (A3)$$

Combining Eqs. (A2) and (A3), we have,

$$d\left(\frac{-pV_0}{T}\right) = Ud\left(\frac{1}{T}\right) + N_e d\left(\frac{-\mu_e}{T}\right). \qquad (A4)$$

In this case, there are two pairs of conjugate variables:

$$\left\{U, \frac{1}{T}\right\}, \left\{N_e, \frac{-\mu_e}{T}\right\}, \qquad (A5)$$

where $U$ and $N_e$ are extensive variables, while $1/T$ and $-\mu_e/T$ are intensive ones. For clarity, we set

$$1/T = \beta$$
$$-\mu_e/T = \gamma$$

Two exact differentials are given by

$$dN_e = L_{11}d\gamma + L_{12}d\beta, \quad (A6)$$

$$dU = L_{21}d\gamma + L_{22}d\beta. \quad (A7)$$

Thus, the combination of Eqs. (A4), (A6) and (A7) leads to a Maxwell relation between $L_{12}$ and $L_{21}$,

$$\Gamma_{21} = \left.\frac{dU}{d\gamma}\right|_\beta = \frac{d^2\left(-\frac{pV_0}{T}\right)}{d\beta d\gamma} = \frac{d^2\left(-\frac{pV_0}{T}\right)}{d\gamma d\beta} = \left.\frac{dN_e}{d\beta}\right|_\gamma = \Gamma_{12}. \quad (A8)$$

For the quasi-static process where $N_e$ is unchanged, we have

$$L_{11}\Delta\gamma + L_{12}\Delta\beta = 0 \Rightarrow 0 = L_{11}\Delta\left(\frac{-\mu_e}{T}\right) + L_{12}\Delta\left(\frac{1}{T}\right). \quad (A9)$$

Referring to Eq. (1), the irreversible Seebeck coefficient is given by,

$$-\frac{\Delta\mu_e}{e\Delta T} = \alpha_r = \frac{1}{eT}\left(\frac{L_{12}}{L_{11}} - \mu_e\right). \quad (A10)$$

Additionally, for the quasi-static process shown with $dT$ vanishing, Eqs. (A6) and (A7) become,

$$e\Delta N_e = eL_{11}\Delta\left(\frac{-\mu_e}{T}\right), \quad (A11)$$

$$\Delta U = L_{21}\Delta\gamma \Rightarrow \Delta U = L_{21}\Delta\left(\frac{-\mu_e}{T}\right). \quad (A12)$$

With $\Delta Q_h = \Delta U - \mu_e \Delta N_e$, we have,

$$\Delta Q_h = L_{21}\Delta\left(\frac{-\mu_e}{T}\right) - \frac{\mu_e}{e}\left(e\Delta N_e\right). \quad (A13)$$

The reversible Peltier coefficient is thus given by

$$\Pi_{\rm r} = \frac{1}{e}\left(\frac{\Gamma_{21}}{\Gamma_{11}} - \mu_{\rm e}\right) = \frac{\Delta Q_{\rm h}}{e\Delta N_{\rm e}}. \qquad (A14)$$

According to Eqs. (A10) and (A14), the reversible reciprocal relation of thermoelectricity is recovered,

$$T\alpha_{\rm r} = \Pi_{\rm r}. \qquad (A15)$$

**Figure Captions**

Figure 1 Schematics for the definition of reversible Seebeck coefficient: A quasi-static process driven by an infinitesimal $\Delta T$ with the iso-electrochemical-potential condition of outer system.

Figure 2 Schematics for the definition of reversible Peltier coefficient: A quasi-static process driven by an infinitesimal $\Delta \mu_e$ with the isothermal condition of outer system.

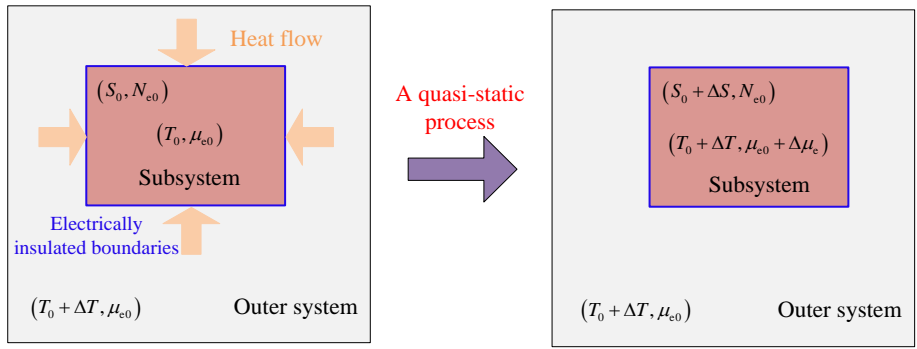

Figure 1 Schematics for the definition of reversible Seebeck coefficient: A quasi-static process driven by an infinitesimal $\Delta T$ with the iso-electrochemical-potential condition of outer system.

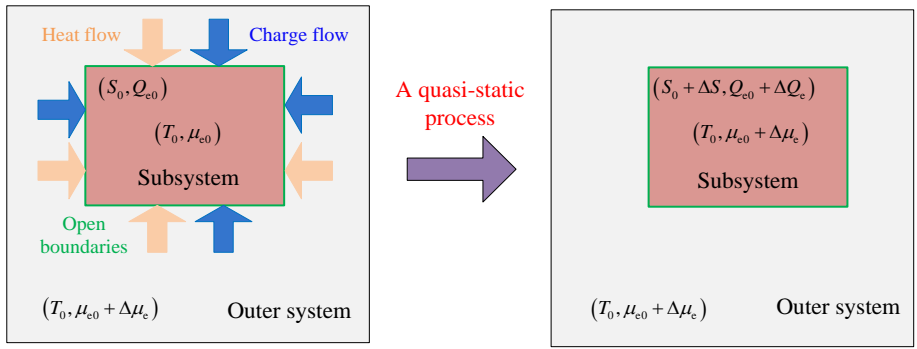

Figure 2 Schematics for the definition of reversible Peltier coefficient: A quasi-static process driven by an infinitesimal $\Delta\mu_e$ with the isothermal condition of outer system.